\documentstyle[12pt,epsf,amstex]{article}
\begin{document}
\setlength{\unitlength}{1mm}
\textwidth 15.0 true cm 
\headheight 0 cm
\headsep 0 cm 
\topmargin 0.4 true in
\oddsidemargin 0.25 true in
\input epsf

\newcommand{\beq}{\begin{equation}}
\newcommand{\eeq}{\end{equation}}
\newcommand{\be}{\begin{eqnarray}}
\newcommand{\ee}{\end{eqnarray}}
\renewcommand{\vec}[1]{{\bf #1}}
\newcommand{\vecg}[1]{\mbox{\boldmath $#1$}}
\renewcommand{\theequation}{\thesection.\arabic{equation}}
\newcommand{\grpicture}[1]
{
    \begin{center}
        \epsfxsize=200pt
        \epsfysize=0pt
        \vspace{-5mm}
        \parbox{\epsfxsize}{\epsffile{#1.eps}}
        \vspace{5mm}
    \end{center}
}

\begin{flushright}

SUBATECH--2002--25

\end{flushright}

\vspace{0.5cm}

\begin{center}

{\Large\bf  Effective Lagrangians for $(0+1)$ and $(1+1)$
dimensionally reduced versions of $D=4$,\ \ ${\cal N} =2$
SYM theory. }

\vspace{1cm}

{\Large A.V. Smilga} \\

\vspace{0.5cm}

{\it SUBATECH, Universit\'e de
Nantes,  4 rue Alfred Kastler, BP 20722, Nantes  44307, France. }
\footnote{On leave of absence from ITEP, Moscow, Russia.}\\
\end{center}

\bigskip

\begin{abstract}
We consider  dimensionally reduced versions of  ${\cal N} =2$\ 
four--dimensi- 
onal supersymmetric Yang-Mills theory
and determine the one-loop effective
Lagrangians associated with the motion over the corresponding
moduli spaces. In the $(0+1)$ case, the effective Lagrangian
describes an ${\cal N}=4$ supesymmetric quantum mechanics of the
Diaconescu--Entin type. In $(1+1)$ dimensions, the effective 
Lagrangian represents a twisted ${\cal N}=4$ supesymmetric
$\sigma$ model due to Gates, Hull, and Ro\^cek. We discuss the
genetic relationship between these two  models and present
the explicit results for all gauge groups.
\end{abstract}

\section{Introduction.}
Perturbative structure of supersymmetric 4--dimensional gauge
theories has been a subject of intense studies since the beginning
of the eighties. Supersymmetry brings about restrictions on the 
perturbative series, and the larger is the supersymmetry, the more 
stringent the restrictions are. Thus, in ${\cal N} =4$ SYM theory
\footnote{To avoid confusion, note that our ${\cal N}$ measures the number
of the irreducible spinor multiplets of supercharges in 4 dimensions
and the number of {\it complex} supercharges in $(0+1)$ and $(1+1)$
dimensions.},
the effective charge is not renormalized at all. In the ${\cal N} 
= 2$ case, only one--loop contribution to the $\beta$ function
survives while all other vanish. In ${\cal N} =1$ theories, all
loops contribute, but there is a rigid relationship between the
$\beta$ function and the anomalous dimensions \cite{beta}.

The perturbative structure of the dimensionally reduced versions
of the original 4--dimensional theories is equally non--trivial
and interesting. In Refs. \cite{star,Ivanov,nov}, we calculated
the effective Lagrangians for the QM versions of 
${\cal N} =1$ supersymmetric electrodynamics and Yang--Mills
theories. The one--loop renormalization of the kinetic term in the
reduced theories involve a power infrared rather than the logarithmic
ultraviolet integral. Still, the corresponding coefficients 
turn out to be rigidly related to the one--loop contribution in
the 4--dimensional $\beta$ function \cite{Akh}. 

 The effective QM Lagrangian represents a nonstandard ${\cal N} =2$
$\sigma$ model: its bosonic part describes the motion over
$3r$--dimensional target space, $r$ being the rank of the group.
(There are certain restrictions for the metric. 
See Ref.\cite{nov} for details.) On the other hand, the effective
$(1+1)$ Lagrangian represents a standard K\"ahlerian $\sigma$-model.
In the simplest case of the ${\cal N} =1$ SQED, the target space
is two--dimensional with the metric 
  \be
\label{metr1+1N1}
ds^2_{1+1} \ =\
 \left[ 1 + \frac {e^2}{4\pi \bar\phi \phi} + \dots \right]
d\bar\phi d\phi\ ,
\ee
where $\phi = L(A_1 + iA_2)/\sqrt{2}$ are the moduli related
to the components of the original 4-dim vector potential $A_\mu$
in two compactified directions.
\footnote{$L = 2\pi R$\ , where $R$ is the radius of compactification. In what follows we set $L=1$. 
Eq. (\ref{metr1+1N1}) is valid when $\bar \phi \phi
\gg e^2$ so that higher loop corrections are suppressed. On the
other hand,  $\bar \phi \phi \ll 1$, otherwise the effects due 
to a finite compactification radius would be important.}
This can be compared with the QM metric
  \be
\label{metr0+1N1}
ds^2_{0+1} \ =\
 \left[ 1 + \frac {e^2}{2 | \vec{A} |^3} + \dots \right]
d\vec{A}^2\ ,
\ee
with 3--dimensional $\vec{A}$.

In the present paper, we address the dimensionally reduced versions
of the $D=4,\ \ {\cal N} =2$ \ theories. As was noticed earlier
\cite{Paban,Akh,2pet} the latter enjoy non--renormalization
theorems that are pretty much similar to their 4--dimensional
counterparts. In particular, the renormalization of the kinetic
term in ${\cal L}_{\rm eff}$ receives contributions only at the
one--loop level, and for ${\cal N} = 4$ theories it is not 
renormalized at all (there are nontrivial contributions to the 
higher-derivative structures, however \cite{sestry}). The exact 
reason why nonrenormalization theorems  in different dimensions are
so similar is not quite clear yet. We think that a better 
understanding of this question could shed also some 
more light on the 
renormalization of supersymmetric theories in 4 dimensions, and 
this our own main motivation for these studies.

 This particular paper, does not address these questions, however,
and  has a technical nature. In the next section, we write some
old and some new formulae referring to the ${\cal N} = 4$ effective
QM theories. They represent nonstandard ${\cal N} = 4$ \ $\sigma$
models living on $5r$--dimensional target space and belong
to the class studied earlier in Ref.\cite{DE}. Sect. 3 is devoted
to the $(1+1)$--dimensional effective theories. They are
${\cal N} = 4$ \ $\sigma$ models.  Somewhat surprisingly, they
are not hyper-K\"ahlerian  but belong to the class of so called
{\it twisted} $\sigma$ models. 
(The latter are not so  widely known, though they
 were first described  back in 1984 
\cite{GHR}.)  

\section{Effective QM Lagrangians}
  As was mentioned above, the models we are looking for belong
to the class of ${\cal N} = 4$ SQM models 
introduced in \cite{DE}. We start 
with recalling the Diaconescu--Entin construction, translating it
into more standard notations and correcting some errors in the
coefficients.

Let us first fix our notations. For 2-component 
complex spinors, the indices are raised and lowered with the
$\epsilon$ symbol (which is nothing but a two- or three-dimensional
 charge conjugation matrix),
 \be
\label{epsconv}
\psi^\alpha = \epsilon^{\alpha\beta} \psi_\beta ,\ \ \ 
 \psi_\alpha = \epsilon_{\alpha\beta} \psi^\beta ,\ \ \ 
\epsilon^{12} = - \epsilon_{12} = 1\ .
  \ee
Now, $\psi_\alpha$ goes over to $\bar\psi^\alpha$ after Hermitian
conjugation. Also, $\psi \chi = \epsilon^{\alpha\beta} \psi_\alpha
\chi_\beta$,\ $\bar\psi \bar\chi = \epsilon_{\alpha\beta} 
\bar\psi^\alpha
\bar\chi^\beta$. 

We will use also the 4--component spinors $\eta_A$ lying in the
fundamental 
representation of $Sp(4)  \equiv SO(5)$. In this case, we prefer
not to distinguish between lower and upper indices and, whenever
necessary, write the charge conjugation matrix $C$ [the symplectic
matrix of $Sp(4)$] explicitly. The latter satisfies the property
$C\gamma_J^T = \gamma_J C$, where $\gamma_J,\ J = 1,\ldots,5$
are Euclidean 5--dimensional $\gamma$ matrices. One of the
possible choices for the latter is
  \be
\label{gammaiC}
\gamma_{1,2,3} = \sigma_{1,2,3} \otimes \sigma_3,\ \ \ 
\gamma_4 = 1\!\!\!1 \otimes \sigma_2,\ \ \ 
\gamma_5 = 1\!\!\!1 \otimes \sigma_1,\ \nonumber \\ 
C = i\sigma_2 \otimes \sigma_1\ =\ 
\left( \begin{array}{cccc} 0&0&0&1 \\   0&0&-1&0 \\   0&1&0&0 \\
-1 & 0 &0 & 0 \end{array} \right)
  \ee
Let us take a standard chiral supervariable 
 \be
\label{Phi}
\Phi = \phi + \sqrt{2} \theta_\alpha \chi^\alpha + \theta^2 F
+ i \dot{\phi} \bar \theta \theta + i \sqrt{2} \theta_\alpha 
\dot{\chi}^\alpha \theta \theta - \frac 12
\ddot{\phi} (\bar \theta \theta)^2 
   \ee
It satisfies the constraint $D_\alpha \Phi = 
[-\partial/\partial \bar\theta^\alpha + i\theta_\alpha 
\partial/\partial t]\Phi \ =\ 0$. 
Consider also a real 3--component
 supervariable \cite{Ivanov}
$$V_k \ =\ -\frac 14  \epsilon^{\beta\gamma} 
(\sigma_k)_\gamma^{\ \alpha} V_{\alpha\beta},\ \ \ \ \ 
V_{\alpha\beta} = V_{\beta\alpha}$$ 
satisfying the constraint
$$D_\alpha V_{\beta\gamma} +
D_\beta V_{\alpha\gamma} +
D_\gamma V_{\alpha\beta} \ =\ 0\ .$$ 
$V_{\alpha\gamma}$ can also be obtained from a generic real
${\cal N} = 2$ supervariable $C$ as 
   \be
  \label{VkviaC}
V_{\alpha\beta} = 
( D_\alpha \bar D_\beta + D_\beta \bar D_\alpha )C \ .
  \ee
$V_k$ represents a supersymmetric generalization of the
vector potential. 
\footnote{A 4--dimensional counterpart of (\ref{VkviaC})
is not  invariant under gauge transformations
$\delta C   = i(\bar \Lambda - \Lambda)$. But the QM
supervariable  (\ref{VkviaC}) is.}
 It is expressed  into components as follows:
  \be
\label{Vcomp}
V_k \ =\ A_k + \bar\psi \sigma_k \theta + \bar\theta
\sigma_k \psi + \epsilon_{kjp} \dot{A}_j \bar\theta 
\sigma_p \theta + D \bar\theta \sigma_k \theta + \nonumber \\ 
i(\bar\theta \sigma_k \dot{\psi} -  
\dot{\bar\psi} 
\sigma_k \theta )\bar\theta \theta + \frac {\ddot{A}_k}4
\theta^2 \bar\theta^2 \ ,
 \ee
where $\psi_\alpha$ is a 2--component spinor and 
$D$ is the auxiliary field.

 Consider the Lagrangian 
   \be
  \label{LK}
\int d^2 \theta d^2 \bar\theta \ {\cal K} (V_k, \bar\Phi, \Phi)
 \ee
with real ${\cal K}$. The Lagrangian (\ref{LK}) has manifest 
${\cal N} =2$ supersymmetry. As was shown in \cite{DE}, for 
a restricted class of functions satisfying the 5--dimensional
harmonicity condition
  \be
\label{harm5}
  \frac {\partial^2{\cal K}}{\partial V_k^2} + 
2 \frac {\partial^2{\cal K}}{\partial \bar\Phi \partial \Phi}
 \ =\ 0\ ,
  \ee
the Lagrangian (\ref{LK}) enjoys the full ${\cal N} = 4$
supersymmetry. To understand this, express (\ref{LK})
into components bearing in mind the condition (\ref{harm5})
  \be
\label{DEcomp}
{\cal L} \ =\ h\left[ \frac 12 \dot{A}_j^2 + \dot{\bar\phi}
\dot{\phi} + \frac i2 (\bar\psi \dot{\psi} - \dot{\bar\psi} \psi
+ \bar\chi \dot{\chi} - \dot{\bar\chi} \chi ) \right]   
+  \nonumber \\ 
\frac 12 (\partial_k h) \epsilon_{kjp} \dot{A}_j [\bar\psi 
\sigma_p \psi + \bar\chi \sigma_p \chi]  
+ \frac i{\sqrt{2}}  \dot{A}_j 
\left[ \frac {\partial h}{\partial \bar\phi}
\bar\psi \sigma_j \bar\chi -
\frac {\partial h}{\partial \phi}
\psi \sigma_j \chi \right] + \nonumber \\ 
\frac i2 \left( \dot{\phi}  
\frac {\partial h}{\partial \phi} - \dot{\bar\phi} 
\frac {\partial h}{\partial \bar\phi} \right) 
(\bar\psi \psi + \bar\chi \chi) + \frac i{\sqrt{2}} 
\partial_k h
\left( \dot{\bar\phi}  \psi \sigma_k \chi - \right. 
\left. \dot{\phi}  \bar\psi \sigma_k \bar\chi \right) - 
\nonumber \\
\frac 1{8h} \left[ \partial_j h (\bar\psi \sigma_j \psi - \bar\chi
\sigma_j \chi) + \sqrt{2} \left( 
\frac {\partial h}{\partial \phi} \psi\chi + 
\frac {\partial h}{\partial \bar\phi} \bar\psi \bar\chi \right)
\right]^2 - \nonumber \\
\frac 1{4h} \left| \psi^2 \frac {\partial h}{\partial \phi}
- \bar\chi^2 \frac {\partial h}{\partial \bar\phi} - 
\sqrt{2} \partial_j h \bar\chi \sigma_j \psi \right|^2 - 
\frac 18 \partial^2 h \left(\bar\psi^2 \psi^2 + \bar\chi^2 \chi^2
\right) -  \nonumber \\ \frac 14 \left( \frac {\partial^2 h}{\partial \phi^2}
\psi^2\chi^2 + \frac {\partial^2 h}{\partial \bar\phi^2}
\bar\psi^2 \bar\chi^2 \right) 
+ \frac 12 \partial_j\partial_k h (\bar\psi \sigma_j \chi) (\bar\chi \sigma_k \psi) + \nonumber \\
 \frac 1{2\sqrt{2}} \left[ \frac {\partial^2 h}
{\partial A_j \partial \phi} \left(\chi^2 \bar\chi \sigma_j \psi 
-\psi^2 \bar\psi \sigma_j \chi  \right)
+ \frac {\partial^2 h}
{\partial A_j \partial \bar\phi}\left(\bar\chi^2 \bar\psi \sigma_j \chi - \bar\psi^2 \bar\chi 
\sigma_j \psi \right)
\right] ,
    \ee
where $h = 2\partial^2 {\cal K}/\partial A_j^2$. 

The bosonic part of Eq.(\ref{DEcomp}) describes the motion
over a 5--dimensional target space with conformally flat
metric 
  \be
\label{5metr}
ds^2 \ =\ h dA_J^2 \ =\ h \left(d\vec{A}^2 + 2 d\bar\phi d\phi
 \right)   
   \ee
($J = 1,\ldots,5;\ \ \phi = (A_4 + iA_5)/\sqrt{2}$. Harmonicity
of ${\cal K}$ implies the harmonicity of $h$:\ $\partial_J
\partial_J h = 0$. The fermion variables $\psi_\alpha$ and
$\chi_\alpha$ enter the expression (\ref{DEcomp}) symmetrically. More exactly,  (\ref{DEcomp}) is invariant under the
transformation 
   \be
   \label{sym5}
\psi \to i\chi, \ \ \ \ \ \ \ \ \  \chi \to i\psi \ .
     \ee
And this
implies that besides manifest supersymmetry transformations
mixing $A$ with $\psi$ and $\phi$ with $\chi$, (\ref{DEcomp})
is also invariant under two other transformations 
  mixing $A$ with $\chi$ and $\phi$ with $\psi$. One can exploit
the symmetry (\ref{sym5}) and define a 
bispinor
 \be
\label{eta}
\eta_A \ =\ \left( \begin{array}{c}  \psi_\alpha \\ 
i\bar\chi_\alpha \end{array} \right) \ .
   \ee
$\eta_A$ belongs to the fundamental representation of $Sp(4) 
\equiv$ spinor representation of $SO(5)$. The Lagrangian (\ref{DEcomp}) can then be rewritten
in $O(5)$ notations as
\footnote{Note the difference in coefficients with Eq.(3.10) of
Ref. \cite{DE} !}
   \be
\label{LO5}
 {\cal L} \ = \ h \left[ \frac 12 \dot{A}^2_J + \frac i2 (\bar\eta
\dot{\eta} - \dot{\bar\eta} \eta ) \right] + \frac i2
\partial_J h \dot{A}_K \ \bar\eta \sigma_{JK} \eta + \nonumber \\
\frac 1{24} \left( 2 \partial_J \partial_K h - \frac 3h \partial_J h
\partial_K h \right) \left( \bar\eta \gamma_J \eta \ 
\bar\eta \gamma_K \eta - \eta C \gamma_J \eta \ 
\bar \eta  \gamma_K C \bar \eta \right)\ ,
   \ee
where $\sigma_{JK} = (1/2) (\gamma_J \gamma_K - \gamma_K
\gamma_J )$. The Lagrangian (\ref{LO5}) 
 is symmetric under charge conjugation
$\eta  \to  C\bar\eta$ [that just coincides with the symmetry (\ref{sym5}) !]. It
enjoys ${\cal N} =4$
supersymmetry for any harmonic function $h(A_J)$. If we also
require the Lagrangian to be $O(5)$ invariant, the function
$h$ must have the form
  \be
 \label{hviaA}
h\ =\ a + \frac c{(A_J^2)^{3/2}}
  \ee
(We will assume that $a = 1$, which can always be achieved by
a proper rescaling. The second term is proportional to 
the Green's function of the Laplacian in
5 dimensions). The corresponding prepotential can be chosen as
   \be
\label{prepot}
{\cal K} \ =\ \frac {R^2}{12} - \frac {\rho^2}8\ -\ 
\frac c{2R} \ln \left(R + \sqrt{R^2 + \rho^2} \right)
\ ,
 \ee
where $R^2 = V_k^2$ and $\rho^2 = 2\bar\Phi \Phi$. Note that
${\cal K}$ need not be and is not $O(5)$ invariant.

It is not so difficult to write a {\it generalized} DE model
including an arbitrary large number of supervariables. Consider
a set of chiral and real 3--vector supervariables
$(V_k^a, \bar\Phi^a, \Phi^a) \equiv V_J^a$,\ $a = 1,\ldots,r$.
The Lagrangian 
  \be
\label{KVa}
   {\cal L} \ =\ \int d^2\theta d^2 \bar\theta \ {\cal K} (V_J^a)
  \ee
is ${\cal N} = 4$ supersymmetric if the function ${\cal K}$ 
satisfies the generalized harmonicity condition
   \be
\label{harm5r}
 \frac {\partial^2 {\cal K}}{\partial V_J^a \partial V_J^b}
\ =\ 0
    \ee
for all $a,b$.

Our goal is to find the QM effective Lagrangians for ${\cal N} =2,
\ \ D=4$ theories. Consider first  Abelian theory. The
moduli space of the ${\cal N} =2$ SQED includes three components
of the vector potential $A_k$ and also a neutral complex
scalar $\phi$. This gives us  5 bosonic degrees of freedom in
${\cal L}_{\rm eff}$. The requirements of $O(5)$ invariance
[the original Lagrangian reduced to (0+1) dimensions had it and
the effective Lagrangian should also have it] and of  
${\cal N} =4$ supersymmetry rigidly determines the form of
${\cal L}_{\rm eff}$. It is given by Eqs.(\ref{LO5}), (\ref{hviaA})
and the only question is the numerical 
value of the coefficient $c$. It is
fixed by the explicit one--loop calculation \cite{Akh,2pet},
   \be
\label{CSQED}
  c_{{\cal N} = 2\ {\rm SQED}} \ =\ 
 c_{{\cal N} = 1\ {\rm SQED}} \ =\ \frac {e^2}2 \ .
   \ee

Now let the original theory be non--Abelian. The moduli space
where the effective Lagrangian lives is now $5r$--dimensional. This
is best seen if treating the QM theory as a result of the 
dimensional reduction of (5+1)--dimensional SYM theory. Classical
vacua correspond to the vanishing field strength $F_{JK} = 0$.
In the QM limit, this implies $[A_J, A_K] = 0$ and hence $A_J$
belongs to the ($r$--dimensional) Cartan subalgebra of the
original Lie algebra. In the simplest $SU(2)$ case,
the rank $r$ is 1, moduli space is 5--dimensional,
and  theory has the same form (\ref{LO5}), (\ref{hviaA}) as in the
Abelian case. Again, the coefficient $c$ is fixed from the one--loop
calculation,
   \be
\label{CSYM}
  c^{SU(2)}_{{\cal N} =2\ \ SYM} \ =\ -g^2 \ .
   \ee
Consider now an arbitrary simple gauge group. The prepotential 
${\cal K}(V^a_k, \bar\Phi^a, \Phi^a)$ can be fixed by the same 
method  as in the ${\cal N} =1$ case \cite{nov}. The effective
Lagrangian is singular when the (positive) root forms
\footnote{In the familiar case of $SU(3)$, there are 3
such root forms corresponding to $T$-spin, $V$-spin, and $U$-spin.
}
$V_k^{(j)} = \alpha_j(V_k^a)$    or $\Phi^{(j)} = \alpha_j(\Phi^a)$
vanish. Indeed, an accurate analysis shows that these are the 
regions in the moduli space where Born--Oppenheimer approximation
breaks down. When, for some particular root $j_0$, 
$| V_k^{(j_0)}|$ and $|\Phi^{(j_0)}|$ 
are small compared to the values
of other root forms, we can forget about the other roots, and
the complicated original Lie algebra is effectively reduced to
$SU(2)$. The effective Lagrangian for the latter was, however,
calculated above.

The result is obtained from the requirement that ${\cal L}_{\rm
eff}$ has a correct limit at small 
$| V_k^{(j_0)}|$ and $|\Phi^{(j_0)}|$ and from the harmonicity 
condition (\ref{harm5r}) with the use of group theory
relations
   \be
  \label{group}
\sum_j \alpha_j(X) \alpha_j(Y) &=& \frac {c_V}2
\sum_{a=1}^r X^a Y^a\ , \nonumber \\
\sum_j c_{jj'}^2 \ =\ d_{j'}  \frac {c_V}2\ ,
     \ee
where $c_V$ is the adjoint Kasimir eigenvalue, 
$c_{jj'} = \langle \alpha_j|\alpha_{j'} \rangle$, 
and $d_j = c_{jj}$ is normalized to 1 for the long roots
and to $1/2$ or $1/3$ (in the case of $G_2$) for the short
roots. The sum in Eq.(\ref{group}) runs over all positive roots.
We obtain
   \be
 \label{prepotr}
F  \ =\  \sum_j \left\{ \frac 1{6c_V} \left[ 
\left(R^{(j)}\right)^2 -
\frac 32 \left(\rho^{(j)}\right)^2 \right] + \right. \nonumber
\\ \left. 
\frac {g^2}{2R^{(j)}} \ln \left[ R^{(j)} + 
\sqrt{\left(R^{(j)}\right)^2 + \left(\rho^{(j)} \right)^2}
 \right] \right\}\ ,
   \ee
where $\left(R^{(j)}\right)^2 = \left(V_k^{(j)}\right)^2$ 
and $\left(\rho^{(j)}\right)^2 = 2 \bar \Phi^{(j)}
\Phi^{(j)}$.

\section{Effective (1+1) Lagrangians.}
Consider first the Abelian case. For the $D=4,\ \ {\cal N} =2$
SQED compactified on $T^2$ 
[or if you will the $D=6,\ \ {\cal N} =1$ SQED compactified
on $T^4$], moduli space is 4--dimensional and can be described
by two complex variables 
  \be
  \label{fisi}
\sigma = (A_1 + iA_2)/\sqrt{2},\ \ \ \ \ \ \ \
\phi = (A_4 + iA_5)/\sqrt{2} 
\ .
  \ee
 One loop calculation brings
about a nontrivial metric in the target space 
($\sigma, \bar\sigma, \phi,  \bar\phi $).
This metric can be related
to the SQM 5--dimensional metric by integrating the latter
over $A_3$ by the same token as the K\"ahlerian metric (\ref{metr1+1N1}) is obtained from the metric (\ref{metr0+1N1}) of the SQM model in the ${\cal N} =1$ case \cite{Akh} ( some further comments about it can be found below),
    \be
\label{metrfisi}
\left. ds^2_{1+1}\right|_{{\cal N} =2} \ =\ 
\left(1 + \delta h_{1+1}\right) [d\bar\sigma d\sigma
+ d\bar\phi d\phi]\ ,  
    \ee
where 
   \be
   \label{relint}
\delta h_{1+1} \ =\ \int_{-\infty}^\infty \frac {dA_3}{2\pi}
\delta h_{0+1} \ = \nonumber \\
 c \int_{-\infty}^\infty \frac {dA_3}{2\pi}
\frac 1{[2\bar\phi \phi + 2\bar\sigma \sigma + A_3^2]^{3/2}}
= \frac c {2\pi(\bar\phi \phi + \bar\sigma \sigma)}\ .
   \ee
We expect the effective action to have the
$\sigma$ model form. One could worry at this point because
the metric (\ref{metrfisi}), (\ref{relint}) is not
hyper-K\"ahlerian (the Ricci tensor and the scalar
curvature do not vanish) while hyper-K\"ahlerian nature of
the metric was shown to be  necessary  for
the standard (1+1) $\sigma$ model to enjoy ${\cal N}= 4$
supersymmetry \cite{Freedman}. In our case, 
${\cal N}= 4$ supersymmetry is there but the metric is not
hyper-K\"ahlerian, and this seems to present a paradox.
The resolution is  that the  $\sigma$ model in hand  
{\it is}  not standard. 

Indeed, the bosonic part of the Lagrangian involves besides the standard kinetic term $h\left( \partial_\alpha \bar \sigma 
\partial_\alpha \sigma   + \partial_\alpha \bar \phi
\partial_\alpha \phi \right) $  also the "twisted" term 
$\propto \epsilon_{\alpha\beta} \partial_\alpha  \sigma
\partial_\beta \phi $ and $\propto \epsilon_{\alpha\beta}
 \partial_\alpha  \bar\sigma
\partial_\beta \bar\phi$. To understand where the twisted term comes from, consider a charged fermion loop in the background
   \be
   \label{backfisi}
   \sigma = \sigma_0 + \sigma_\tau \tau + \sigma_z z,\ \ \ \ 
\phi = \phi_0 + \phi_\tau \tau + \phi_z z
  \ee
  ($\tau$ is the Euclidean time). The contribution to the 
effective action is $\propto \ln \det \| {\mathfrak D}\|$, 
where ${\mathfrak D}$ is the 6--dimensional Euclidean Dirac 
operator, which can be written in the form
   \be
   \label{Dir6} 
   {\mathfrak D}  \ =\ i \frac{\partial}{\partial\tau} + \gamma_3
   \frac{\partial}{\partial z} -i (\gamma_1 A_1 + \gamma_2 A_2 +
   \gamma_4 A_4 + \gamma_5 A_5 )
     \ee
 [$\gamma$ matrices are defined in Eq.(\ref{gammaiC}) and 
$A_{1,2,4,5}$  in Eq.(\ref{fisi})].  
     
 Now, if $A_4$ and $A_5$ were absent, we could write 
 ${\mathfrak D} = \gamma_4 (i\tilde{\gamma}_\mu {\cal D}_\mu )$ , with
 $$\mu = 1,2,3,4;\  {\cal D}_4 = \frac{\partial}{\partial\tau} ,\ 
 {\cal D}_3 = \frac{\partial}{\partial z},\ {\cal D}_{1,2}=  - i A_{1,2}; \  \tilde{\gamma_4} = \gamma_4, \tilde{\gamma}_{1,2,3} = -i\gamma_4 \gamma_{1,2,3} $$
and then use the squaring trick
  \be
  \label{square}
  \det \| {\mathfrak D} \| \ =\ \det \| i \tilde{\gamma}_\mu {\cal D}_\mu \| \ =\ 
\det\,\!^{1/2} \left\| - {\cal D}^2 + \frac i2 
\tilde{\sigma}_{\mu\nu} F_{\mu\nu} \right\|\ ,
   \ee
with $F_{14} = \partial A_1 /\partial \tau$, etc. The effective action
would be proportional to 
    \be
       \label{intFF}
{\rm Tr} \{\sigma_{\mu\nu} \sigma_{\alpha\beta} \}
F_{\mu\nu} F_{\alpha\beta} \int \frac {d^2p}{4\pi^2} 
\frac 1{(p^2 + 2\bar\sigma \sigma )^2}\ ,
  \ee
  which gives the renormalization of the kinetic term while the twisted term does not appear. The squaring trick works also in the case where $A_{4,5}$ are nonzero, but do not depend on $\tau,z$. Then $2\bar\phi \phi$ is just added to $-{\cal D}^2$ in Eq.(\ref{square}) and to $2\bar\sigma \sigma$ in Eq.(\ref{intFF}) leading to Eq.(\ref{relint}). But in the generic case the fermion determinant cannot be reduced to $\det^{1/2} \| - {\cal D}^2 + \frac i2 \sigma_{\mu\nu} F_{\mu\nu} \|$. The basic reason for this impasse is that one cannot adequately ``serve" six components of the gradient with only five $\gamma$ matrices.
\footnote{By the same reason, the squaring trick does not work for Weyl 2--component fermions in 4 dimensions: three Pauli matrices that are available in that case are not enough to do the job.}
  As a result, the extra twisted term in the determinant appears.
  
  We need not perform an explicit calculation here as the twisted and all other terms in the Lagrangian are fixed by supersymmetry. The twisted ${\cal N} = 4$ supersymmetric $\sigma$ model was constructed almost 20 years ago\cite{GHR}. At that time it did not attract much attention. Recently, there is some revival of interest in the GHR model: it happened to pop up in some  string--related problems \cite{brany,DiaSei}. It also pops up as the effective $(1+1)$ Lagrangian in the case under study.

It was shown that, for ${\cal N} = 4$ supersymmetric generalization to be possible, the conformal factor in the metric $h(\bar\sigma, \sigma, \bar\phi, \phi)$ should satisfy the harmonicity condition
  \be
  \label{harm4}
 \frac {\partial^2 h}{\partial \bar\sigma \partial \sigma } +
 \frac {\partial^2 h}{\partial \bar\phi \partial \phi} \ =\ 0\ .
  \ee
  Obviously, (\ref{relint}) satisfies it everywhere besides the origin. The relationship of (\ref{harm4}) to the 5--dimensional harmonicity condition for the metric in the effective SQM model (\ref{LO5}) is also obvious. Indeed, integrating a $D$--dimensional harmonic function over one of the coordinates like in (\ref{relint}), we always arrive at a $(D-1)$--dimensional harmonic function.

To construct the full action, consider along with the standard chiral multiplet $\Phi$ satisfying the conditions ${\cal D}_\alpha \Phi= 0$ also a {\it twisted} chiral multiplet $\Sigma$ which satisfies the constraints
$$ {\cal D}^1\Sigma  \equiv \bar{\cal D}_+ \Sigma = 0,\ \ \ \ \ \ 
{\cal D}_2 \Sigma \equiv {\cal D}_- \Sigma = 0\ .$$
Now, $\Phi$ and $\Sigma$ are expressed into components as follows
    \be
    \label{Phicomp}
  \Phi \ =\ \phi + \sqrt{2} (\theta_+ \chi_- - \theta_- \chi_+) +
  i(\partial_+ \phi) \bar\theta_+ \theta_+ +
   i(\partial_- \phi) \bar\theta_- \theta_-  -
   \nonumber \\
  i\sqrt{2} \left[ \bar\theta_+ (\partial_+ \chi_+) + \bar\theta_-(\partial_- \chi_-) \right] \theta_+ \theta_- - ( \partial_+ \partial_- \phi)  \bar\theta_+ \theta_+ \bar\theta_- \theta_-  + 2 \theta_+ \theta_-  F\ ,
   \ee
   and 
     \be
    \label{Sigcomp}
  \Sigma\ =\ \sigma + \sqrt{2} (\bar\theta_+ \psi_- - \theta_- \bar\psi_+) -
  i(\partial_+ \sigma) \bar\theta_+ \theta_+ +
   i(\partial_- \sigma) \bar\theta_- \theta_-    +
   \nonumber \\
  i\sqrt{2} \left[ \theta_+ (\partial_+ \bar\psi_+)  + \bar\theta_- \partial_- \psi_- \right] \theta_- \bar\theta_+ + ( \partial_+ \partial_- \sigma)  \bar\theta_+ \theta_+ \bar\theta_- \theta_-  + 2 \bar\theta_+ \theta_-  G\ ,
   \ee     
 where $\partial_\pm = \partial_t \pm \partial_z$. 
 The twisted multiplet (\ref{Sigcomp}) is closely related to the multiplet (\ref{Vcomp}). Actually, the QM version of Eq.(\ref{Sigcomp}) 
(obtained when the spatial derivatives are suppressed, $\partial_\pm
\to \partial_t$) just coincides with $(V_1 + iV_2)/\sqrt{2}$
(and $G = (D + i \dot A_3)/\sqrt{2}$ ). 
 
 We see that the twisted multiplet differs from the standard one by a pure convention: $\Sigma$ is obtained from $\Phi$ by interchanging
$\theta_+$ and $\bar\theta_+$. This means that the change $\Phi \to \Sigma$ in any standard action involving $\Phi$ would change nothing. However, one can write nontrivial Lagrangians involving {\it both} $\Phi$ and $\Sigma$. The twisted $\sigma$ model is determined by the expression
  \be
  \label{LGHR}
  {\cal L} \ =\ \int d^2 \theta d^2 \bar\theta 
  \ {\cal K} (\bar\Phi, \Phi; \bar\Sigma, \Sigma)\ ,
     \ee
     where the prepotential ${\cal K}$ satisfies the harmonicity condition,
  \be
  \label{harm4K}
 \frac {\partial^2 {\cal K}}{\partial \bar\Sigma \partial \Sigma } +
 \frac {\partial^2 {\cal K}}{\partial \bar\Phi \partial \Phi} \ =\ 0\ .
  \ee
The condition (\ref{harm4K}) is required if we want the theory to be ${\cal N} = 4$ supersymmetric. Only for a harmonic ${\cal K}$, the fermion interchange symmetry (\ref{sym5}) holds for the fermion kinetic term
 \be
  \label{fermkin}
{\cal L}_{\rm kin}^{\rm ferm}\ =\  {ih} \left[ \bar\chi_+ \partial_+ \chi_+ +   \bar\chi_- \partial_- \chi_- +  \bar\psi_+ \partial_+ \psi_+ +   \bar\psi_- \partial_- \psi_-  \right]\ ,
    \ee
($h = 4\partial^2 {\cal K}/\partial \bar\sigma \partial \sigma 
    = - 4\partial^2 {\cal K}/\partial \bar\phi\partial \phi $)   and for the  
full Lagrangian (cf. Eq.(6.11) of Ref. \cite{Gates} )
   \be
   \label{4chlena}
   {\cal L} \ =\ {\cal L}_{\rm kin}^{\rm bos} + {\cal L}_{\rm kin}^{\rm ferm} + {\cal L}_{\rm mixed} + {\cal L}_{\rm 4f}\ ,
  \ee
  where
  \be
  \label{boskin}
 {\cal L}_{\rm kin}^{\rm bos}  \ =\ 
 h \left[ \left| \partial_\alpha \phi \right|^2 + 
\left| \partial_\alpha \sigma \right|^2 \right] \ + \nonumber \\
 4  \left[\frac{\partial^2 {\cal K}} {\partial \sigma 
\partial \phi } 
\epsilon_{\alpha\beta}(\partial_\alpha \sigma)
(\partial_\beta \phi) +
\frac{\partial^2 {\cal K}}
{ \partial \bar\sigma \partial \bar\phi } \epsilon_{\alpha\beta} 
(\partial_\alpha \bar \sigma)(\partial_\beta \bar\phi)  
      \right]  
\ ,
    \ee
    \be
    \label{mixed}
   {\cal L}_{\rm mixed} =
   \frac i2 (\bar\psi_+ \psi_+ + \bar\chi_+ \chi_+ )\left(
   \frac {\partial h}{\partial  \bar\sigma } \partial_+ \bar\sigma + 
   \frac {\partial h}{\partial \phi } \partial_+ \phi  - 
   \frac {\partial h}{\partial \sigma } \partial_+ \sigma - 
   \frac {\partial h}{\partial \bar\phi } \partial_+ \bar\phi \right) 
        \nonumber \\
- \frac i2 (\bar\psi_- \psi_- + \bar\chi_- \chi_- )\left(
   \frac {\partial h}{\partial \bar\sigma } \partial_- \bar\sigma -
   \frac {\partial h}{\partial \phi } \partial_- \phi  - 
   \frac {\partial h}{\partial \sigma } \partial_- \sigma  +
   \frac {\partial h}{\partial \bar\phi } \partial_- \bar\phi \right)   +
   \nonumber \\
   - i \psi_+ \chi_+ \left( \frac {\partial h}{\partial \bar\sigma }
   \partial_+ \bar\phi  - \frac {\partial h}{\partial \phi }
   \partial_+ \sigma \right) -
    i \bar\psi_+ \bar\chi_+ \left( \frac {\partial h}{\partial \sigma }
   \partial_+\phi  - \frac {\partial h}{\partial \bar\phi }
   \partial_+ \bar\sigma \right) - \nonumber \\
   i \psi_- \chi_- \left( \frac {\partial h}{\partial \phi }
   \partial_- \bar\sigma  - \frac {\partial h}{\partial \sigma }
   \partial_- \bar\phi \right) -
    i \bar\psi_- \bar\chi_- \left( \frac {\partial h}{\partial \bar\phi }
   \partial_-\sigma  - \frac {\partial h}{\partial \bar\sigma }
   \partial_- \phi \right) \ ,
       \ee 
   \be
   \label{4ferm}
 {\cal L}_{\rm 4f}  = - \frac{\partial^2 h}{\partial \bar\sigma 
\partial \sigma}  (\bar\psi_+ \psi_+ + \bar\chi_+ \chi_+ )
(\bar\psi_- \psi_- + \bar\chi_- \chi_- ) + 
\frac{\partial^2 h}{\partial \bar\sigma^2}  \bar\psi_- \bar\chi_- 
\psi_+
\chi_+    \nonumber \\ -
\frac{\partial^2 h}{\partial \phi^2}  \psi_+ \psi_- \chi_+ 
\chi_-  + \frac{\partial^2 h}{\partial \sigma^2}  \bar\psi_+ 
\bar\chi_+ \psi_-  \chi_-  - 
\frac{\partial^2 h}{\partial \bar\phi^2}  \bar\psi_+ \bar\psi_- 
\bar\chi_+  \bar\chi_-    \nonumber \\ +
 (\bar\psi_+ \psi_+ + \bar\chi_+ \chi_+ ) 
 \left( \frac {\partial^2 h }{\partial
 \bar\sigma \partial \bar\phi } \bar\psi_- \bar\chi_- 
 -  \frac {\partial^2 h }{\partial \sigma \partial \phi }
 \psi_- \chi_-  \right) + 
\nonumber \\ +
(\bar\psi_- \psi_- + \bar\chi_- \chi_- ) 
 \left(   \frac {\partial^2 h }{\partial
 \bar\sigma \partial  \phi } \psi_+ \chi_+ 
 -
  \frac {\partial^2 h }{\partial 
\sigma \partial \bar\phi } \bar\psi_+  \bar\chi_+  
\right)  
\nonumber \\ -
 \frac 1h 
 \left|  \frac {\partial h}{\partial \bar\sigma } \bar\chi_- \psi_+  -
   \frac {\partial h}{\partial \phi }  \psi_+ \psi_- +
   \frac {\partial h}{\partial \sigma } \bar\chi_+ \psi_- -
   \frac {\partial h}{\partial \bar\phi } \bar\chi_+
   \bar\chi_-  \right|^2   \nonumber \\ -
   \frac 1h 
 \left|  \frac {\partial h}{\partial \bar\sigma } \bar\chi_- \chi_+
  +   \frac {\partial h}{\partial \phi }   \psi_- \chi_+  -
   \frac {\partial h}{\partial  \sigma } \bar\psi_+ \psi_- -
   \frac {\partial h}{\partial \bar\phi } \bar\chi_- 
   \bar\psi_+  \right|^2 \ .
   \ee    
 By the same token as in the DE model discussed in the previous section, the symmetry (\ref{sym5}) brings about   
two extra  supersymmetries mixing $\sigma$ with $\chi$ and $\phi$ with $\psi$ on top of two    manifest 
supersymmetries mixing $\sigma$ with $\psi$ and $\phi$ with $\chi$.

The metric $h$ was fixed in Eqs.(\ref{relint}), (\ref{CSQED}), (\ref{CSYM}). There is a freedom in the choice of the prepotential: two functions ${\cal K}$ and ${\cal K}'$ related as
  \be
  \label{KKprim}
  {\cal K}' \ =\ {\cal K} + f(\bar\sigma, \phi) + \bar f(\sigma, \bar\phi) + g(\sigma, \phi) + \bar g(\bar\sigma, \bar\phi)
   \ee
   define one and the same theory (adding $f + \bar f$ leaves ${\cal L}$ invariant while adding $g + \bar g$ changes it by a total derivative). One of the possible choices for ${\cal K}$ is 
\cite{Hitchin,DiaSei}
   \be
   \label{Kres}
   {\cal K} \ =\ \frac {\bar\Sigma \Sigma - \bar\Phi \Phi}4 +
   \frac c{8\pi} \left[ F\left( \frac {\bar\Sigma \Sigma }{\bar\Phi \Phi} \right)  - \ln \Phi \ln \bar\Phi \right]\ ,
 \ee
 where
   \be
   \label{Spence}
   F(\eta) \ =\ \int_1^\eta \frac {\ln(1+\xi)}\xi \ d\xi
    \ee
 is the Spence function. Substituting it in ${\cal L}^{\rm bos}_{\rm kin}$, we obtain 
   \be
  \label{Lbosres}
 {\cal L}_{\rm kin}^{\rm bos}  \ =\ 
 \left[ 1 + \frac c{2\pi(\bar\sigma \sigma + \bar\phi \phi)} \right]
  \left[ \left| \partial_\alpha \phi \right|^2 + \left| \partial_\alpha \sigma \right|^2 \right] \  \nonumber \\
 - \frac c{2\pi(\bar\sigma \sigma + \bar\phi \phi)} 
 \left[ \frac
 {\sigma}{\bar\phi} \epsilon_{\alpha\beta} (\partial_\alpha \bar\sigma)(\partial_\beta \bar\phi)  +\frac  {\bar\sigma} \phi 
 \epsilon_{\alpha\beta}(\partial_\alpha \sigma)(\partial_\beta \phi) 
 \right]  
\ .
    \ee
The twisted term is a  2--form $F$. Its external derivative $dF$
can be associated with the torsion (the freedom (\ref{KKprim}) 
of choice of
${\cal K}$ corresponds to adding to $F$ the external derivative
of the 1-form ``organized'' from the functions 
$f,\bar f, g, \bar g$. The torsion is invariant under such a 
change.) 
Now, $F$  is self-dual, 
$F = F^*$ (with the convention $\sigma = (x+iy)/\sqrt{2}, \ 
\phi = (z + it)/\sqrt{2}$). One can observe that the ``action" $\int F \land F^*$ diverges logarithmically.
\footnote{One can wonder whether a variant of $\sigma$ model with {\it non-Abelian} self-dual torsion exist ?..}

It is clear that the Lagrangians (\ref{LGHR}), (\ref{4chlena}) on one hand and (\ref{LK}), (\ref{LO5}) on the other hand are closely related, like the superfields (\ref{Sigcomp}) and (\ref{Vcomp}) are. 
Of course, (\ref{LO5}) is not obtained from  (\ref{4chlena})  by a trivial dimensional reduction: the degrees of freedom counting is different, etc. The relationship is established in the same way as in the ${\cal N} =1$ case \cite{Akh} : one should take the
{\it functional integral} with the Lagrangian (\ref{LO5}) and perform the integration over $\prod_t dA_3(t) $. After that the metric is transformed as in Eq.(\ref{relint}), the terms involving the derivatives with respect to $A_3$ disappear and the 4--fermion term is transformed as
   \be
    \label{4fnov}
{\cal L}_{4f} \ \to \ {\cal L}_{4f} + \frac {(\partial_J h)(\partial_K h) }{8h} \ \bar\eta \sigma_{J3} \eta \ \bar\eta \sigma_{K3} \eta \ .
   \ee
   The QM Lagrangian thus obtained  coincides
 with the Lagrangian (\ref{4chlena}) where all spatial 
derivatives (and thereby the twisted bosonic term) are suppressed, 
$\partial_\pm \to \partial_t$. 

Consider now a generic non--Abelian case. For a simple Lie group of 
rank $r$, the effective Lagrangian is
\be
  \label{Lr}
  {\cal L} \ =\ \int d^2 \theta d^2 \bar\theta 
  \ {\cal K} (\bar\Phi^a, \Phi^a; \bar\Sigma^a, \Sigma^a)\ ,
     \ee
where $a = 1,\ldots,r$ and the expression for ${\cal K}$ is derived exactly in the same
way as for the SQM model of the previous section [see Eq.(\ref{prepotr})]. We have
  \be
   \label{Kr}
   {\cal K} \ =\ \sum_j \left\{ \frac 1{2c_V} 
\left[\bar\Sigma^{(j)} \Sigma^{(j)} - \bar\Phi^{(j)} \Phi^{(j)} \right] \right.
\nonumber \\ 
\left. -   \frac {g^2}{8\pi} \left[ F \left( \frac {\bar\Sigma^{(j)} \Sigma^{(j)} }
{\bar\Phi^{(j)} \Phi^{(j)}} \right)  - \ln \Phi^{(j)} \ln \bar\Phi^{(j)}
 \right]
\right\}
\ ,
 \ee
where $\Sigma^{(j)} = \alpha_j(\Sigma^a)$, etc. The prepotential (\ref{Kr}) 
satisfies a generalized harmonicity condition
   \be
  \label{harm4Kr}
 \frac {\partial^2 {\cal K}}{\partial \bar\Sigma^a \partial \Sigma^b } +
 \frac {\partial^2 {\cal K}}{\partial \bar\Phi^a \partial \Phi^b} \ =\ 0
  \ee
for all $a,b$.

  \section{Discussion}
The results obtained in this paper are closely parallel to the well-known results
derived earlier in Refs.\cite{SW,D3}. In \cite{SW} the effective Lagrangian for the 
4D ${\cal N} =2$ SYM theory was constructed. It involved $r$ different Abelian gauge
fields $V^a$, the moduli complex variables $\phi^a$, and their superpartners. 
The Lagrangian is {\it exact} as far as the quadratic in derivatives terms are concerned.
The same program was partly 
carried out for the  ${\cal N} =2$ SYM theory
with one spatial dimension compactified \cite{D3}. The Lagrangian represents a 
complicated
hyper--K\"ahlerian $\sigma$ model. Again, it is exact when higher derivative terms
are disregarded.

We have solved here the same problem, but for the theories with two and three spatial
dimensions compactified. Our results are exact in the same sense as above. As we have
shown, the effective $(0+1)$ and $(1+1)$ models are closely related. It would be
interesting to explore their relationship to $(3+1)$ and $(2+1)$ models in more
details. But one difference is already seen. The effective Seiberg--Witten Lagrangian
not only takes into account the one loop renormalization of the effective charge (at this
level, the relationship was explored back in \cite{Akh}), but also sums up nontrivial
multi-instanton effects. No trace of these nonperturbative effects is left in $(1+1)$ and
in $(0+1)$ dimensions.

\section*{Acknowledgements}
We are indebted to E.T.  Akhmedov and K.G. Selivanov
for illuminating discussions and to S.J. Gates, Jr, 
for useful correspondence.

\end{document}